\documentclass[12pt]{article}

\usepackage{amsmath}
\usepackage{amsfonts}
\usepackage{amssymb}
\usepackage{mathrsfs}
\usepackage{multicol}
\usepackage{color}
\usepackage{graphicx}
\usepackage{pstricks}
\usepackage{pst-node}
\usepackage{wrapfig}
\usepackage{enumerate}

\usepackage{cite}

\usepackage{amsfonts,amssymb}
\usepackage{amsthm}
\usepackage{amscd}
\definecolor{gray1}{gray}{0.1}
\definecolor{gray2}{gray}{0.2}
\definecolor{gray3}{gray}{0.3}
\definecolor{gray4}{gray}{0.4}
\definecolor{gray5}{gray}{0.5}
\definecolor{gray6}{gray}{0.6}
\definecolor{gray7}{gray}{0.7}
\definecolor{gray8}{gray}{0.8}
\definecolor{gray9}{gray}{0.9}

\newpsobject{showgrid}{psgrid}{subgriddiv=1,griddots=10,gridlabels=6pt}

\definecolor{dark-green}{rgb}{0,0.7,0}
\definecolor{dark-blue}{rgb}{0,0.2,0.5}
\definecolor{med-blue}{rgb}{0,0.7,1}
\definecolor{mblue}{rgb}{0,0.2,1}
\definecolor{cnc}{rgb}{0.8,0,0}
\definecolor{light-red}{rgb}{1,0.8,0.8}
\definecolor{dark-yellow}{rgb}{1,0.8,0}
\definecolor{light-blue}{rgb}{0.8,0.9,1}
\definecolor{verylight-blue}{rgb}{0.93,0.95,1}
\definecolor{light-yellow}{rgb}{1,0.9,0.8}
\definecolor{grey}{gray}{0.88}

\newpsobject{showgrid}{psgrid}{subgriddiv=1,griddots=10,gridlabels=6pt}

\begin{document}

\thispagestyle{empty}

\setlength{\abovecaptionskip}{10pt}

\begin{center}
{\Large\bfseries\sffamily{Solitons and black hole in shift symmetric scalar-tensor gravity \\
with cosmological constant}}
\end{center}
\vskip 3mm
\begin{center}
{\bfseries{\sffamily{Yves Brihaye$^{\rm 1}$, Betti Hartmann$^{\rm 2}$, 
Jon Urrestilla$^{\rm 3}$}}}\\
\vskip 3mm{$^{\rm 1}$\normalsize Physique-Math\'ematique, Universit\'e de Mons-Hainaut, 7000 Mons, Belgium}\\
{$^{\rm 2}$\normalsize{Instituto de F\'isica de S\~ao Carlos (IFSC), Universidade de S\~ao Paulo (USP), CP 369, \\
13560-970 , S\~ao Carlos, SP, Brazil}}\\
{$^{\rm 3}$\normalsize{Department of Theoretical Physics, University of the Basque Country UPV/EHU,
48080 Bilbao, Spain}}
\end{center}

\begin{abstract} 
We demonstrate the existence of static, spherically symmetric {\it globally regular}, i.e.  solitonic solutions of a 
shift-symmetric scalar-tensor gravity model with negative cosmological constant. The norm of the
Noether current associated to the shift symmetry is finite in the full space-time. 
We also discuss the corresponding black hole
solutions and demonstrate that the interplay between the scalar-tensor coupling and the cosmological constant leads to the existence of new branches of solutions. 
To linear order in the scalar-tensor coupling, the asymptotic space-time corresponds to an Anti-de Sitter space-time with a non-trivial scalar field on its conformal boundary. This allows 
the interpretation of our solutions in the context of the AdS/CFT correspondence. 
Finally, we demonstrate that -- for physically relevant, small values of the scalar-tensor coupling -- solutions with positive cosmological constant do not exist in our model.

\end{abstract}

\section{Introduction}
Einstein's theory of General Relativity (GR) is up to the present day the best theory we have to describe the gravitational interaction. In recent detections of gravitational waves from colliding black hole binaries by the LIGO collaboration \cite{Abbott:2016blz, Abbott:2016nmj, Abbott:2017vtc, Abbott:2017oio, Abbott:2017gyy}  
no deviations from GR have been found. Next to the direct observation  of gravitational
waves (one of the predictions of the theory), these observations have also conclusively shown that
black holes do, indeed, exist in the universe.  Accepting this, however, leads to a problem~:
black hole solutions of standard GR are plagued with physical, i.e. space-time singularities at their gravitational
center. These are, in general, not observable for an outside observer due to the existence of an event horizon,
but it shows that this classical theory of gravity possesses limits. In particular, up to today it seems
impossible to reconcile Quantum physics, which is the fundamental basis of the Standard Model of Particle physics,
with GR. One of the best candidates is String Theory, which contains the assumed mediator of the gravitational interaction, the graviton,
naturally. Since the energy scale at which Quantum Gravity (QG) theories should become relevant are out of reach for present day accelerators,
there are different ways to test whether extensions of GR are necessary.

One such possibility is to observe strong gravity events such as the collision of two black holes or the recently observed collision of two neutron stars 
using both gravitational waves as well as Gamma-rays \cite{Monitor:2017mdv, TheLIGOScientific:2017qsa,Troja:2017nqp}. This allows to test gravity theories to a very high precision with the help of multi-messenger gravitational wave (GW) observations. 

Another way to test gravity theories is to study theoretical predictions that are also detectable at low(er) energies.
One such prediction of String Theory is the Anti-de Sitter/Conformal Field Theory (AdS/CFT) correspondence \cite{adscft,ggdual} that
states that a gravity theory in a $d$-dimensional AdS space-time possesses exactly the same number of
degrees of freedom as the $(d-1)$--dimensional CFT on the conformal boundary of AdS. 
This duality is a weak--strong coupling correspondence and has mainly been used in the context of
classical $d$-dimensional GR in AdS describing a strongly coupled  CFT on the $(d-1)$--dimensional boundary with the extra  dimension giving the 
renormalization group (RG) flow. One such application is the description of high-temperature superconductors in terms of holographic duals \cite{hhh,reviews}, another one the application to heavy ion collisions and the quark-gluon plasma (see \cite{Ammon:2015wua} and references therein).

In recent years, Horndeski gravity models \cite{horndeski} have gained a lot of attention. These models include, in general, higher curvature invariants and non-linear
couplings between scalar, vector and tensor fields. One such example are scalar-tensor gravity models in which a scalar field, the so-called galileon, is non-minimally
coupled to the tensor part \cite{Nicolis:2008in,Deffayet:2009wt,Deffayet:2011gz,Kobayashi:2011nu,Deffayet:2013lga,Hui:2012qt, Charmousis:2014mia, Babichev:2016rlq}. These models contain a shift symmetry of the scalar field that leads to the existence of a conserved Noether current. Consequently, black hole solutions have been studied in these models \cite{Sotiriou:2013qea, Sotiriou:2014pfa}, which, however,
have a diverging norm of the Noether current on the horizon. In \cite{Babichev:2017ab} solutions with vanishing Noether current have been constructed.
Interestingly, in these same models it has been conjectured that globally regular, star-like solutions in asymptotically flat space-time 
should not be able to carry Galilean ``hair", while it has been suggested that the no-go theorem for astrophysically relevant solutions could be avoided in asymptotically
de Sitter (dS) space-time \cite{Lehebel:2017fag}.

In this paper, we discuss solutions of a Galileon scalar-tensor gravity model with a linear coupling between the scalar field and the Gauss-Bonnet term
including a cosmological constant. Recent multi-messenger GW astronomy that puts strong bounds on the GW speed, disfavoures these models
\cite{171005901} in their original version without cosmological constant. In fact, originally these
models had been studied in order to solve the dark energy problem through the inclusion
of a dynamical scalar field which renders the cosmological solutions self-accelerating.  In this paper, we will show that the explicit presence of a positive cosmological constant
does neither allow black hole nor star-like, solitonic solutions for relevant, small values of the
scalar-tensor coupling. On the other hand, we will demonstrate that the presence of a negative cosmological
constant in the model allows for solitonic solutions as well as new branches of black hole solutions
as compared to the case with vanishing cosmological constant. Hence, while the motivation to study these
solutions is no longer given in an astrophysical and/or cosmological context, we suggest that the dual description
of strongly coupled phenomena within the AdS/CFT correspondence  could lead to interesting new models in the future.

Our paper is organized as follows~: in Section 2, we give the model and equations, while in Section 3 and 4 we discuss soliton and black hole solutions,
respectively. Section 5 contains our summary and conclusions.

\section{The model}
The model we are studying in this paper is a Horndeski scalar-tensor model which possesses a shift symmetry in the scalar field 
$\phi\rightarrow \phi + a_{\mu} x^{\mu} + c$, where $a_{\mu}$ is a constant co--vector and $c$ is a constant.  Its action
reads~: 
\begin{equation}
\label{action}
S =  \int  {\rm d}^4x  \sqrt{-g} \left[R-2\Lambda  +  \frac{\gamma}{2}  \phi {\cal G} -  
\partial_{\mu} \phi  \partial^{\mu} \phi   \right] \ ,
\end{equation}
where the Gauss-Bonnet term ${\cal G}$ is given by
\begin{equation}
 {\cal G} = R^{\mu \nu \rho \sigma} R_{\mu \nu \rho \sigma} - 4 R^{\mu \nu}R_{\mu \nu} + R^2  \ .
\end{equation}
$\gamma$ is the scalar-tensor coupling and $\Lambda \neq 0$ is the cosmological constant 
\footnote{Note that the action can be supplemented by suitable boundary and counter terms  in order to render the action finite. For the discussion in the following these are, however, not necessary.}.
Units are chosen such that $16\pi G\equiv 1$ and the scalar-tensor coupling $\gamma$ is related to the $\alpha$ used in \cite{Sotiriou:2014pfa}
by $\gamma=4\alpha$. For $\Lambda=0$, this model and its black hole
solutions have been studied in \cite{Sotiriou:2013qea,Sotiriou:2014pfa,Babichev:2017ab}, while it was recently pointed out that solitonic, i.e. star-like
solutions that are globally regular, do not exist in this model \cite{Lehebel:2017fag}. In the following, we will 
demonstrate that this does not hold true for the model with $\Lambda < 0$. 

Varying the  action (\ref{action}) with respect to the metric and the scalar field gives the following gravity equations~:
\begin{equation}
\label{eom1}
G_{\mu\nu} +\Lambda g_{\mu\nu} - \partial_{\mu} \phi \partial_{\nu} \phi + \frac{1}{2}g_{\mu\nu} \partial_{\alpha} \phi \partial^{\alpha} \phi + \frac{\gamma}{2}
{\cal K}_{\mu\nu} = 0 \ ,
\end{equation}
where 
\begin{equation}
       {\cal K}_{\mu\nu} =( g_{\rho\mu} g_{\sigma\nu} + g_{\rho\nu} g_{\sigma\mu})\nabla_{\lambda} (\partial_{\gamma} \phi \epsilon^{\gamma \sigma\alpha\beta} \epsilon^{\delta\eta } R_{\delta\eta\alpha\beta}) \ ,
\end{equation}
as well as the scalar field equation
\begin{equation}
\label{eom2} 
\square \phi = - \frac{\gamma}{2} {\cal G} \  .
\end{equation}
We are interested in 
spherically symmetric and static solutions and hence choose for the Ansatz~:
\begin{equation}
\label{metric}
ds^2=-N(r) \sigma(r)^2 dt^2 + \frac{1}{N(r)} dr^2  + r^2 \left(d\theta^2 + \sin^2 \theta d \varphi^2\right)  \ , \   \phi=\phi(r)  \ .
\end{equation}
Inserting this Ansatz into the equations of motion (\ref{eom1}) and (\ref{eom2}) results in a coupled system of non-linear ordinary differential equations that has to be solved
subject to the appropriate boundary conditions. The explicit form of the gravity equations (\ref{eom1}) reads~:
\begin{equation}
\label{em1}
 4\gamma (1-N)\phi'' - 2\gamma(3N-1) \phi' \frac{N'}{N} + r^2 \phi'^2 +2r\frac{N'}{N} - \frac{2}{N} +2 + 2\Lambda \frac{r^2}{N}  = 0  \ ,
\end{equation}
\begin{equation}
\label{em2}
 \frac{1}{N^2} + 3\gamma \phi' \left(2\frac{\sigma'}{\sigma} + \frac{N'}{N}\right) - \left[ 1+ (r + \gamma\phi')\left(2\frac{\sigma'}{\sigma} 
 + \frac{N'}{N}\right)   - \frac{1}{2}r^2 \phi'^2\right] \frac{1}{N} - \Lambda \frac{r^2}{N} = 0 \ ,
\end{equation}
\begin{eqnarray}
\label{em3}
{\cal F}'\left(2\gamma \phi'- \frac{r}{N}\right) + {\cal F} \left[2\gamma \phi''+ 2\gamma\phi'\left(2\frac{N'}{N} + \frac{\sigma'}{\sigma} \right) - \frac{r}{N}\left(\frac{N'}{N} + \frac{\sigma'}{\sigma} \right) - \frac{1}{N}\right] - \frac{r}{N} \phi'^2 - \frac{N'}{N^2} - 2 r\frac{\Lambda}{N^2} = 0 \  \ \ , \ \ 
 \end{eqnarray}
 where ${\cal F}= \frac{N'}{N} + 2\frac{\sigma'}{\sigma}$  and the prime now and in the following denotes the derivative with respect to $r$.  Finally, the scalar field equation (\ref{eom2})  is~:
 \begin{eqnarray}
 \label{em4}
  \gamma\left[(N-1)\left( 4\frac{\sigma''}{\sigma} + 2\frac{N''}{N} + 6 \frac{\sigma' N'}{\sigma N} + \frac{N'}{N}\right) +
4 \frac{\sigma' N}{\sigma N}
+ 2\frac{N'^2}{N}\right] + 2r^2 \left[\phi'' 
+  
\left(\frac{\sigma'}{\sigma}+\frac{N'}{N} + \frac{2}{r}\right)\phi' \right]
  = 0 \ .
  \end{eqnarray}
Note that the system of equations does not depend on $\phi(r)$ explicitly, but only on $\phi'(r)$.
\\
The model contains a shift symmetry $\phi\rightarrow \phi + a_{\mu} x^{\mu} + c$, which leads
to the existence of a locally conserved Noether current $J^r$ with norm $(J_r J^r)^{1/2}$ that results from the invariance of the kinetic term
under this transformation and the fact that the Gauss-Bonnet term is a total divergence in 4-dimensional space-time, respectively. 
In our choice of coordinates the norm reads \cite{Babichev:2017ab}~:
\begin{equation}
\label{current}
\left(J_r J^r\right)^{1/2} = \frac{\gamma(N-1)}{2r^2}\left(\frac{N'}{N} + \frac{2\sigma'} {\sigma}\right) - \phi' \ .
\end{equation}
For $\gamma=0$ the system of equations corresponds to Einstein gravity including a cosmological constant. For $\gamma > 0$,
the asymptotically non-vanishing curvature of the space-time sources the scalar field -- even at $r\rightarrow \infty$. It is, thus, not surprising that the solution to the equations of motion does not behave like
a ``pure" (Anti-) de Sitter ((A)dS) space-time asymptotically. For $r\rightarrow \infty$, we find the following behaviour
for the metric functions and the scalar field derivative~:
\begin{equation}
\label{expansion_infinity}
N(r) = C_1 - \frac{\lambda}{3} r^2 - \frac{M}{r} + {\rm O}(r^{-2}) \ \ , \ \ \sigma(r)=1 + {\rm O}(r^{ -2}) \ \ , \ \
\phi'(r) = -\frac{C_2}{r} + {\rm O}(r^{-3})  
\end{equation}
where ${\lambda}$, $C_1$ and $C_2$ fulfil
\begin{equation}
{\lambda} \left(3 + \frac{10}{9} \gamma^2 {\lambda}^2\right) = 3\Lambda \ \ \ , \ \ \   C_1 =
\frac{16 \gamma^4\left(\frac{\lambda}{3}\right)^4 + 12\gamma^2\left(\frac{\lambda}{3}\right)^2 + 1}{40\gamma^4\left(\frac{\lambda}{3}\right)^4 + 14\gamma^2\left(\frac{\lambda}{3}\right)^2 + 1}
 \ \ \ \ , \ \ \ \
C_2 = -\frac{2}{3}\gamma{\lambda}  \ 
\end{equation}
and the parameter $M$ determines the gravitational mass  of the solution (see \cite{charmousis_AdS} for a summary and references therein)~:
\begin{equation}
M_{\rm grav} = \frac{8\pi}{16\pi G} M = 8\pi  M \ ,
\end{equation}
where in the last equality we have used our convention $16\pi G =1$. 

The relation between $\lambda$ and $\Lambda$ implies that the signature of these constants is equal, i.e. negative (positive) $\Lambda$ implies negative (positive) $\lambda$, and
that $\vert\lambda\vert < \vert \Lambda\vert$ for $\gamma\neq 0$ with $\vert\lambda\vert$ a decreasing function with $\Lambda$ fixed and $\gamma$ increasing.
To state it differently, for a fixed value of $\Lambda$, the asymptotic space-time deviates increasingly from (A)dS for increasing $\gamma$.

However, for relevant, small values of $\gamma$ our solutions are -- to linear order in $\gamma$ -- global (A)dS with a scalar field $\psi(r):=\phi'(r)$ that falls of linearly. 
Since the system depends only on $\psi(r)$, it is possible to interpret the asymptotically AdS solution in terms of the AdS/CFT correspondence
by letting the coefficient
$C_2=\frac{2}{3}\gamma{\lambda}\approx\frac{2}{3}\gamma\Lambda$ give the dual description of the expectation value of a quantum operator of dimension unity in the CFT on the conformal boundary. On the other hand, the solution with $\Lambda > 0$ possesses a horizon at $r_c=\sqrt{3/\lambda}$ which -- to linear order 
in $\gamma$ -- is equal to the cosmological horizon of dS space-time. 
However, as we will show below, neither soliton nor black hole solutions with (approximately) dS asymptotics exist, though the expansion (\ref{expansion_infinity})
is valid for both signs of $\Lambda$.

\section{Solitonic solutions}
As pointed out in \cite{Lehebel:2017fag}, star-like, globally regular and asymptotically flat 
solutions with non-trivial scalar field do not exist in the model we are studying here. However, when $\Lambda <  0$, these solutions exist, as we will show in the following.
Let us start with the behaviour of the functions  close to $r=0$. We find~:
\begin{equation}
\label{ex_zero}
N(r) = 1 + \frac{n_2}{2} r^2 + {\rm O}(r^3) \ \ , \ \ \sigma(r)=\sigma_0\left(1 + \frac{\sigma_2}{2} r^2 + {\rm O}(r^3)\right) \ \ , \ \ \phi(r)=\frac{\phi_2}{2} r^2 + {\rm O}(r^3) \ \ ,
\end{equation}
where $n_2$, $\sigma_2$ and $\phi_2$ fulfil the following relations~:
\begin{equation}
n_2 = \frac{2 \Lambda}{6\gamma \phi_2 - 3} \ \ , \ \ \sigma_2= \frac{\gamma \Lambda \phi_2}{3(1-2\gamma\phi_2)^2} 
\end{equation}
and
\begin{equation}
8\gamma^3 \phi_2^4 - 12\gamma^2 \phi_2^4 + 6\gamma \phi_2^2 + \left(\frac{2}{3} \gamma^2 \Lambda^2 -1 \right) \phi_2 - \frac{2}{9} \gamma \Lambda^2  = 0  \ . 
\end{equation}
$\sigma_0\equiv \sigma(0)$ is a constant to be determined numerically (see below for numerical results).
This expansion already implies that both $\gamma\neq 0$ and $\Lambda \neq 0$ are necessary
to obtain solitonic solutions. 
In order to make this more evident, let us look at the expansion in $\gamma$ -- we would expect $\gamma$ in any realistic model to be small and hence the expansion in $\gamma$ can give hints on how the
existence of the soliton manifests itself. 
We find~:
\begin{equation}
\label{soliton_gamma}
N(r)=1 - \frac{\Lambda}{3} r^2 + \gamma^2 N_2(r) + {\rm O}(\gamma^4) \ \ \ , \ \ \ 
\sigma(r) = 1 + \gamma^2 \Sigma_2(r) + {\rm O}(\gamma^4) \ \ , \ \  \phi'(r) = \gamma \varphi_1(r) + {\rm O}(\gamma^3) \ ,
\end{equation}
where the function $N_2(r)$, $\Sigma_2(r)$ and $\varphi_1(r)$ are given by~:
\begin{equation}
    N_2(r) = - \frac{2}{9} \Lambda^2 + \frac{10}{81} \Lambda^3 r^2 + \frac{2}{r} \left(-\frac{\Lambda^3}{27}\right)^{1/2} {\rm arctan}\left(r \left(-\frac{\Lambda}{3}\right)^{1/2}\right) \ \ , 
\end{equation}
and 
\begin{equation}
\Sigma_2(r) = \frac{\Lambda^2}{27(\Lambda r^2 -3)}   \ \ \ , \ \ \  \varphi_1(r) = \frac{2 r \Lambda^2}{3(\Lambda r^2 -3)}  \ .
\end{equation}
We conclude that only for $\gamma\neq 0$ and $\Lambda < 0$ we can have globally regular solutions. 
For $\Lambda > 0$, the function $\varphi_1$ tends to infinity at the cosmological horizon $r_c=\sqrt{3/\Lambda}$. Hence, {\it globally regular dS solutions
do not exist in our model}, at least not for physically relevant, small values of $\gamma$. 
\\
It is also important to note that the norm of the Noether current (\ref{current}) fulfils
$\sqrt{J_r J^r}(r=0)=0$ and that it is finite in the full interval of the radial coordinate $r$ for $\Lambda < 0$ (see numerical results). Moreover, we can only define the gravitational mass of the soliton
at quadratic order in $\gamma^2$. The mass parameter $M$ (see (\ref{expansion_infinity})) can be found by considering the $\gamma$ expansion for $r\rightarrow \infty$ and is given by
\begin{equation}
M= - \pi\gamma^2 \left(-\frac{\Lambda^3}{27}\right)^{1/2} + {\rm O}(\gamma^4)   \ .
\end{equation}
For $\gamma=0$, $\Lambda < 0$ the solution corresponds to global AdS and consequently has mass equal to zero, while the solitonic
solutions for $\gamma > 0$ possesses negative mass for $\Lambda < 0$.

\subsection{Numerical construction}
\begin{figure}
\begin{center}
{\includegraphics[width=10cm]{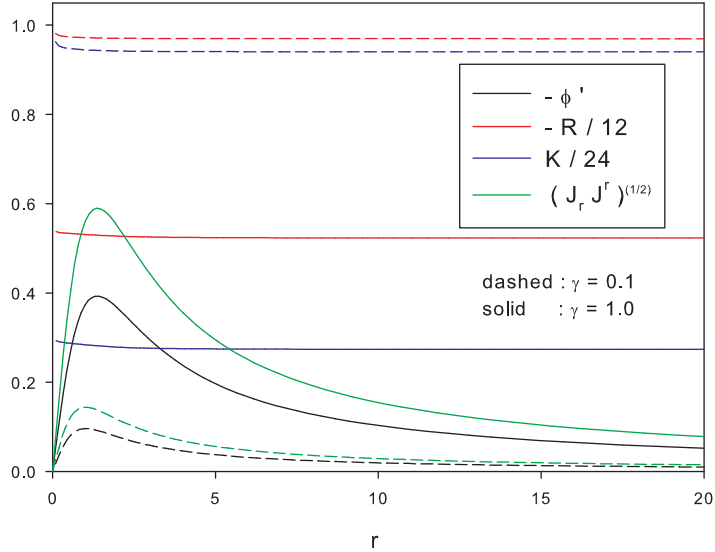}}
\caption{We show the dependence of $-\phi'$, the Ricci scalar $R$, the Kretschmann scalar
$K=R_{\nu\mu\rho\sigma} R^{\nu\mu\rho\sigma}$ and the norm of the Noether current $(J_r J^r)^{1/2}$ on the radial coordinate $r$ for
soliton solutions with $\Lambda=-3$ and $\gamma=0.1$ (dashed) and $\gamma=1.0$ (solid), respectively.   
\label{fig:invariants}}
\end{center}
\end{figure}

\begin{figure}
\begin{center}
\includegraphics[width=10cm]{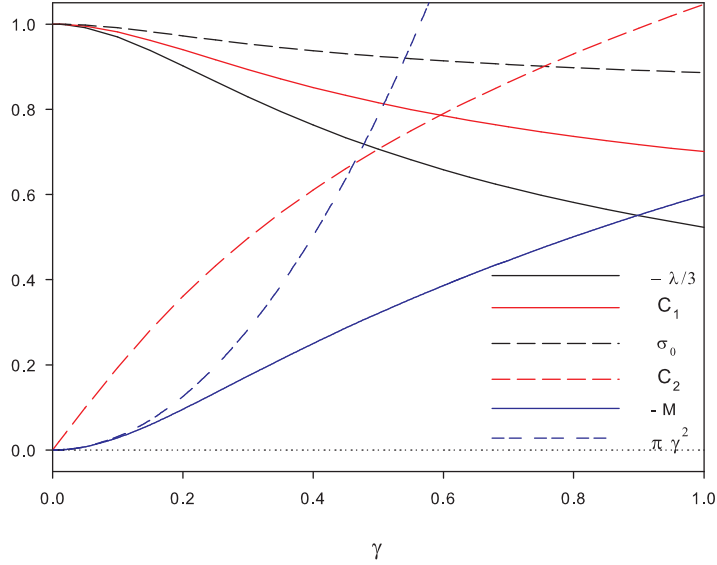}
\caption{We show the dependence of the coefficients appearing in the asymptotic expansion (see (\ref{expansion_infinity})) as well as $\sigma_0$ (see (\ref{ex_zero})) on $\gamma$ for soliton
solutions with $\Lambda=-3$. Note that the negative value of the mass parameter $-M$, which for our choice of $\Lambda=-3$ reads
$-M=\pi \gamma^2 + O(\gamma^4)$, fits the curve $\pi\gamma^2$ very well for small values of $\gamma$.
\label{fig:soliton_data}}
\end{center}
\end{figure}

In the following, we will choose $\Lambda$ negative since our analysis above has shown that globally regular dS solutions do not
exist. In order to solve the equations (\ref{em1}) -- (\ref{em4}) numerically, suitable combinations can be
made and the system reduces to two equations of first order (the equations for $N$ and $\sigma$)
as well as one equation of second order (the equation for $\phi$). 
We will hence fix 4 boundary conditions in the following \footnote{Since the equations of motion depend only on $\phi'$, the 
system is effectively a system of three 1st order equations and $\phi(0)$ is a free parameter,
which can be chosen to be equal to zero due to the shift symmetry in the model.}. Using the results of the expansion close to $r=0$, we choose the following conditions for globally regular solutions~:
\begin{equation}
N(0)=1 \ \ , \ \phi(0)=0 \ \ , \ \ \ \phi'(0)=0 \ \ \ \ , \ \ \  a(r\rightarrow \infty)\rightarrow 1 \ .
\end{equation}
In Fig.\ref{fig:invariants} we show $\phi'(r)$ for $\Lambda=-3$ and two values of $\gamma$ together with
the Kretschmann scalar $K= R_{\nu\mu\rho\sigma} R^{\nu\mu\rho\sigma}$, the Ricci scalar $R$ and the norm of
the Noether current. 
As can be seen, the space-time is perfectly regular everywhere, contains no physical singularities and possesses a finite norm of the Noether current.
Moreover, $\phi'$ tends asymptotically to zero  and increases in value over the whole range of $r$ when increasing $\gamma$.  The norm of the Noether current is zero at $r=0$, increases to a maximal value close to the center of
the soliton and then falls off to zero asymptotically. With the increase of $\gamma$, the norm increases, but
stays qualitatively the same.
The Kretschmann scalar $K$ and the Ricci scalar $R$ deviate increasingly
from their ``pure" AdS values $K_{\rm AdS}=-8\Lambda=24$ and $R_{\rm AdS}=4\Lambda=-12$ for increasing $\gamma$. This is nothing else but the statement
made above that the space-time is asymptotically AdS only to linear order in $\gamma$. 
This becomes also clear by virtue of Fig.~\ref{fig:soliton_data}, in which we give the
dependence of the coefficients appearing in the asymptotic expansion (see (\ref{expansion_infinity})) as well
as the value of $\sigma_0$  (see (\ref{ex_zero})) on $\gamma$ for $\Lambda=-3$. Again, for the allowed range of the parameter $\gamma$, the value of $\sigma_0$ stays perfectly finite. The mass parameter $M$ is zero for $\gamma=0$, which corresponds to global AdS, and becomes
increasingly negative when increasing $\gamma$ for $\Lambda$ fixed. We have hence found a continuous
branch of solitonic solutions that is directly connected to global AdS.

\section{Black hole solutions}
For $\gamma=0$, our model has a black hole solution, namely the Schwarzschild-AdS solution (SAdS) (see e.g. \cite{charmousis_AdS} and references therein) with
\begin{equation}
N(r) = 1 - \frac{\Lambda}{3} r^ 2 - \frac{M}{r}  \ ,  \ \sigma(r)\equiv 1 \ \ , \ \ \phi(r)\equiv 0  \ ,
\end{equation}
and the mass parameter $M$ is related to the horizon $r_h$ by $M=r_h - \frac{\Lambda}{3} r_h^3$. 

In order to understand the deformation of this black hole solution in the presence of the scalar field, we have first studied the expansion of the solution in powers of $\gamma$. 
We find~:
\begin{equation}
      N(r) = 1 - \frac{\Lambda}{3}  r^2 - \frac{M}{r} + \gamma^2 \tilde{N}_2(r) + {\rm O}(\gamma^4) \ \ , \ \ 
      \sigma(r) = 1 + \gamma^2 \tilde{\Sigma}_2(r) + {\rm O}(\gamma^4) \ \ , \ \ 
      \phi'(r) = \gamma \Phi_1(r) + {\rm O}(\gamma^3)  \ ,
\end{equation}
and function $\Phi_1(r)$ reads~: 
 \begin{equation}
      \Phi_1(r) = \frac{\left(r^2+r r_h + r_h^2\right)\left(\frac{2}{9}r^3 r_h \Lambda^2 + 
      \frac{1}{9}r_h^4 \Lambda^2 - \frac{2}{3} r_h^2 \Lambda + 1 \right)}{r^4 r_h\left(\frac{\Lambda}{3}r^2  + \frac{\Lambda}{3} r r_h  + \frac{\Lambda}{3}r_h^2  - 1\right)} \ .
 \end{equation}
 In particular, we find from this expression the dominant asymptotic term $\Phi_1(r\rightarrow \infty) \sim \frac{2}{3} \frac{\Lambda}{r}$,  which is in excellent agreement
with our numerical construction (see below).
The expressions for $\tilde{N}_2(r)$ and $\tilde{\Sigma}_2(r)$ are very lengthy, that is why we do not present them here. Let us just note that, 
asymptotically, the expansion in $\gamma$ is equivalent to the $\gamma$ expansion of the soliton solution, see (\ref{soliton_gamma}).

The temperature of a static black hole is given in terms of its surface gravity $\kappa$ evaluated at the horizon $r=r_h$~:
\begin{equation}
\label{T_bh}
T_H=\frac{\kappa}{2\pi}\left\vert_{r=r_h} \right. \ \ \ , \ \ \  \kappa^2=-\frac{1}{4} g^{tt} g^{ij} \partial_i g_{tt} \partial_j g_{tt} \vert_{r=r_h} \ \ , \ \ i,j=1,2,3 \ .
\end{equation}
For our Ansatz (\ref{metric}) and using (\ref{em1}) we find~:
\begin{equation}
4\pi T_H=\left(N' \sigma\right)_{r=r_h} = \frac{1-\Lambda r_h^2}{r_h+\gamma \phi'\vert_{r=r_h}} \sigma (r_h)\ ,
\end{equation}
where $\phi'\vert_{r=r_h}$ can be expressed in terms of $\Lambda$, $r_h$ and $\gamma$ (see  (\ref{phip})  below).

\begin{figure}
\begin{center}
\includegraphics[width=10cm]{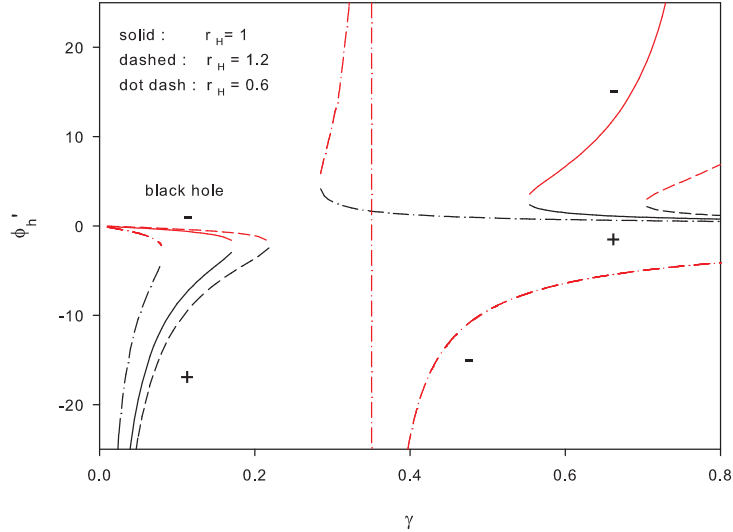}
\caption{We show the domain of existence of black hole solutions with scalar hair and $\Lambda=-3$ in the 
$\phi'_h$--$\gamma$--
plane. The curves denote the critical value of $\gamma$ (see (\ref{gammacr})) for the four different branches (see also (\ref{phip})) for $r_h=1.2$ (dashed),
$r_h=1.0$ (solid) and $r_h=0.6$ (dotted-dashed). Note that the $+$ and $-$ indicate the two different
branches as given by (\ref{phip}). Moreover, the vertical dotted-dashed line is the value of $\gamma$ at which
the denominator of (\ref{phip}) becomes zero for $r_h=0.6$. 
\label{fig:domain}}
\end{center}
\end{figure}

\subsection{Numerical construction}

In order to construct black hole solutions numerically, we choose -- like in the soliton case -- $a(r\rightarrow \infty)\rightarrow 1$, while
we now have to impose boundary conditions on the regular (non-extremal) horizon $r=r_h$. These conditions read~:
\begin{equation}
\label{bc_rh}
      N(r_h) = 0 \ \ , \ \ \phi(r_h)=0 \ \ 
\end{equation}
as well as the following condition for the scalar field derivative $\phi'(r)$~: 
\begin{equation}
\label{phip}
\phi'\vert_{r=r_h}=\frac{\pm \sqrt{\Delta}\vert \Lambda r_h^2 -1 \vert + 2\Lambda \gamma^2 r_h^3 
-6 \Lambda\gamma^2 r_h  + \Lambda r_h^5 - r_h^3}{2\gamma(2\Lambda \gamma^2  - \Lambda  r_h^4  + r_h^2)}
\end{equation}
with
\begin{equation}
\label{delta}
\Delta=4 \gamma^4\Lambda^2 r_h^2-24 \gamma^4\Lambda+8 \gamma^2\Lambda r_h^4-12\gamma^2 r_h^2+r_h^6  \ .
\end{equation}
The requirement $\Delta \geq 0$ gives the intervals in $\gamma$ for which black holes with
regular horizon and non-trivial scalar hair exist. We find~:
(assuming that $\gamma \ge 0$)~:
\begin{equation}
\label{gammacr}
\gamma_{\rm cr}(\Lambda)^{(\pm)}= \left(\frac{\pm\sqrt{3} r_h^2\sqrt{\Lambda^2 r_h^4 -2 \Lambda r_h^2 +3}  - 2\Lambda r_h^4 + 3 r_h^2}{2\Lambda(\Lambda r_h^2 -6)}\right)^{1/2} \ .
\end{equation}
Solutions with regular horizon then exist in the intervals $\gamma\in [0:\gamma_{\rm cr}^{(-)}]$ and
for $\gamma\geq \gamma_{\rm cr}^{(+)}$.  Moreover, the denominator in (\ref{phip}) can become
zero leading to the divergence of $\phi'_h$. Hence depending on the choices of $\Lambda$ and $r_h$ there
might be one value of $\gamma$, where solutions with regular horizon do not exist. 

\begin{figure}
\begin{center}
\includegraphics[width=10cm]{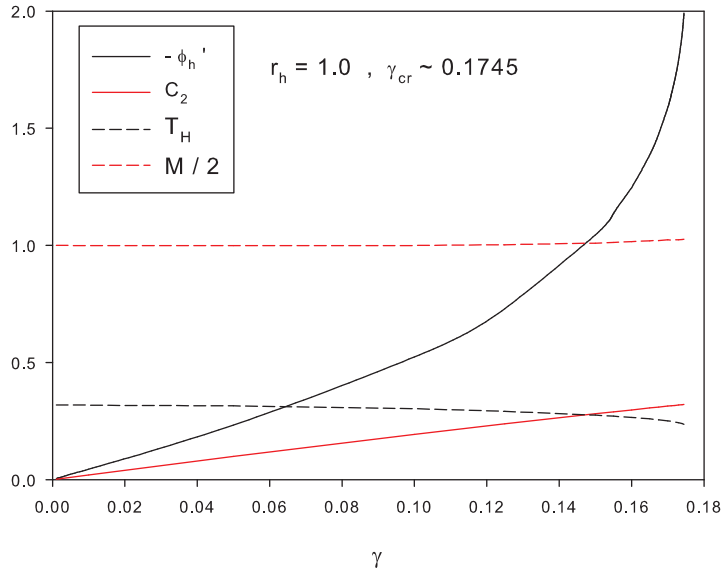}
\caption{We show the temperature $T_{\rm H}$, the  constant $C_2$ (see (\ref{expansion_infinity})), 
$-\phi'_h$ (see (\ref{phip})) as well as the mass parameter $ M$ in dependence of $\gamma$ for $\Lambda=-3$ and $r_h=1$. Note that in this
case $\gamma_{\rm cr}\approx 0.1745$.   
\label{fig_bh}}
\end{center}
\end{figure}

Equations (\ref{phip}) and (\ref{gammacr}) indicate that we should expect to have four branches for $\Lambda \neq 0$. This is shown in Fig.\ref{fig:domain} for $\Lambda=-3$
and different values of $r_h$~: the curves in this plot show the value of $\phi'\vert_{r=r_h}\equiv \phi'_h$ in dependence of $\gamma_{cr}$. The plus and minus signs
indicate the different branches as given by (\ref{phip}), while the different intervals in $\gamma$ are obvious from this plot.  Note that for $\Lambda=0$ only the branches close to $\gamma=0$ exist. Hence, the non-vanishing cosmological
constant leads to the existence of new branches that are disconnected from the $\gamma=0$, $\Lambda=0$ limit. For decreasing $r_h$ the value of $\gamma_{cr}$ decreases
such that for $r_h\rightarrow 0$, the branches disappear. On the other hand, the extend in $\gamma$ of the two disconnected branches
increases with decreasing $r_h$. One could then speculate that the limit $r_h\rightarrow 0$ exists, corresponding to a soliton solution. Although we have shown above that 
soliton solutions actually exist in our model for $\Lambda < 0$, we find, however, that the branches never reach the corresponding soliton solution which has $\phi'_h=0$.
To state it differently~: the soliton solutions presented above do not arise in a smooth limit taking $r_h\rightarrow 0$ for the black hole solutions.

We have also constructed the solutions on the branches that exist for large enough values of $\gamma$ and find that these have -- in general --
$\phi'\rightarrow \infty$ for some intermediate $\tilde{r}$ with $\tilde{r} \in [r_h:\infty[$. This is why we do not discuss them further here.

In Fig.~\ref{fig_bh} we show the dependence of the black hole temperature,  the  constant $C_2$ (see (\ref{expansion_infinity})), the value
$-\phi'_h$ (see (\ref{phip})) as well as the mass parameter $M$ in dependence of $\gamma$ for $\Lambda=-3$ and $r_h=1$.
In this case, the interval in which solutions exist is $\gamma\in [0:\gamma_{\rm cr}^{(-)}] $ with $\gamma_{\rm cr}^{(-)}\approx 0.174$ and for $\gamma \gtrsim 0.550$ with the
denominator of (\ref{phip}) diverging at $\gamma=2/3\approx 0.667$. 
The temperature of the black hole $T_{\rm H}$ decreases from its  ``pure" 
AdS value $T_{\rm H, AdS}=
(-\Lambda r_h + r_h^{-1})/(4\pi) \approx 0.3183$ at $\gamma=0$ when increasing $\gamma$, while the coefficient
$C_2$ increases. If we interpret $C_2$ as on order parameter, we find that $C_2$ increases with decreasing temperature $T_{\rm H}$, a phenomenon typically observed in superconductors.  
For $\gamma_{\rm cr}^{(-)} \approx 0.174$ the derivative of the scalar function at the horizon diverges, which makes
the black hole temperature $T_{\rm H}$ by virtue of (\ref{T_bh}) tend to zero as $\gamma\rightarrow\gamma_{\rm cr}^{(-)}$, while $C_2$ stays finite. For $\gamma \in [0.174:0.550]$ no globally regular
black hole solutions exist, while for  $\gamma \geq 0.550$ and $\gamma \neq 0.667$ black holes with non-trivial scalar field and regular horizon at $r=r_h$ exist, but as mentioned above, these solutions become singular at a finite value of $r > r_h$. 
Finally, the mass parameter $M$ does not depend strongly on $\gamma$ and stays close to its SAdS
value $M=r_h-\frac{\Lambda}{3} r_h^3 \equiv 2$ for  our choice of parameters $r_h=1$, $\Lambda=-3$.

\section{Conclusions}
In this paper, we have presented evidence that the non-existence theorem for globally regular solutions
of shift-symmetric scalar-tensor gravity models does not extend to the case with negative cosmological constant. We have constructed
globally regular, solitonic solutions that have AdS asymptotics to linear order in $\gamma$. The corresponding black hole
solutions possess a regular horizon at $r=r_h$, but do not tend to the soliton solutions in the limit $r_h\rightarrow 0$.
We also observe that the presence of the negative cosmological constant allows new branches of black hole solutions, which, however, 
possess a diverging scalar field derivative at finite distance outside the horizon. 
As mentioned above, it will be interesting to understand the application of our solutions in the context of the AdS/CFT correspondence,
e.g. in the holographic description of high-temperature superconductors. The parameter $\gamma$ triggers the existence of
a non-trivial scalar field in the space-time and for black holes we find a typical superconductor behaviour, namely
the order parameter increases with decreasing temperature.  In the standard approach to holographic
superconductors \cite{hhh,reviews}, the scalar field is minimally coupled to an electromagnetic field (and a tensor gravitational field if backreaction is taken into account) and this coupling triggers
the spontaneous formation of scalar hair below a critical temperature of the planar AdS black hole.
For our model it will be interesting to see how the existence of a non-trivial scalar field on the conformal boundary can be interpreted, especially in the
context of the existence of ``gaps'' in $\gamma$, where regular black holes with scalar hair do not exist. This is currently under investigation.

Our results also indicate that neither solitonic nor black hole solutions in a space-time with positive cosmological constant exist, and,
we have, in fact, not been able to construct the solutions with the appropriate boundary conditions. We believe that this is
related to the fact that the system of ordinary differential equations of the model is effectively a system of three 1st order equations
and, consequently, only 3 boundary conditions are not trivial. Since, however, for the construction of black hole solutions, we would need
to fix the metric function $N(r)$ to be equal to zero on both the regular horizon $r_h$ as well as the cosmological horizon $r_c$ and require
in addition the scalar field to be regular at these two points, the number of boundary conditions needed appears too large for the system of
equations. 

\vspace{1cm}

{\bf Acknowledgements} 
 BH would like to thank FAPESP for financial support under
grant number {\it 2016/12605-2} and CNPq for financial support under
{\it Bolsa de Produtividade Grant 304100/2015-3}.  



 \end{document}